\documentstyle[amssymb,floats,aps,prb,epsf]{revtex}
\epsfclipon
\begin{document}
\twocolumn[\hsize\textwidth\columnwidth\hsize\csname
@twocolumnfalse\endcsname

\draft
\title{Electronic structure of kinetic energy driven
superconductors}
\author{Huaiming Guo and Shiping Feng}
\address{Department of Physics, Beijing Normal University, Beijing
100875, China}
\maketitle
\begin{abstract}
Within the framework of the kinetic energy driven
superconductivity, we study the electronic structure of cuprate
superconductors. It is shown that the spectral weight of the
electron spectrum in the antinodal point of the Brillouin zone
decreases as the temperature is increased. With increasing the
doping concentration, this spectral weigh increases, while the
position of the sharp superconducting quasiparticle peak moves to
the Fermi energy. In analogy to the normal-state case, the
superconducting quasiparticles around the antinodal point disperse
very weakly with momentum. Our results also show that the striking
behavior of the superconducting coherence of the quasiparticle
peaks is intriguingly related to the strong coupling between the
superconducting quasiparticles and collective magnetic
excitations.
\end{abstract}

]

\bigskip

\narrowtext

The parent compounds of cuprate superconductors are the Mott
insulators with an antiferromagnetic (AF) long-range order
(AFLRO), then changing the carrier concentration by ionic
substitution or increasing the oxygen content turns these
compounds into the superconducting (SC)-state leaving the AF
short-range correlation still intact \cite{bedell,kastner}. The
single common feature of cuprate superconductors is the presence
of the two-dimensional CuO$_{2}$ plane \cite{bedell,kastner}, and
it seems evident that the unusual behaviors of cuprate
superconductors are dominated by this CuO$_{2}$ plane
\cite{anderson}. This layered crystal structure leads to that
cuprates superconductors are highly anisotropic materials, then
the electron spectral function $A({\bf k},\omega)$ is dependent on
the in-plane momentum \cite{shen,eschrig,fink}. Experimentally, an
agreement has emerged that at least in the SC-state, the
electronic quasiparticle excitations are well defined and are the
entities participating in the SC pairing
\cite{shen,eschrig,fink,campuzano,dlfeng,campuzano1}. According to
a comparison of the density of states as measured by scanning
tunnelling microscopy \cite{dewilde} and angle-resolved
photoemission spectroscopy (ARPES) spectral function
\cite{shen,ding} at the antinodal point, i.e., the $[\pi,0]$ point
of the Brillouin zone, on identical samples, it has been shown
that there is the presence of a shallow extended saddle point in
the $[\pi,0]$ point \cite{shen,eschrig,fink}, where the d-wave SC
gap function is maximal, then the most contributions of the
electron spectral function come from the $[\pi,0]$ point
\cite{shen,eschrig,fink,ding}. Moreover, recent improvements in
the resolution of ARPES experiments allowed for an experimental
verification of the particle-hole coherence in the SC-state and
Bogoliubov-quasiparticle nature of the sharp SC quasiparticle peak
near the $[\pi,0]$ point \cite{matsui,campuzano}. It is striking
that in spite of the high temperature SC mechanism and observed
exotic magnetic scattering \cite{yamada,dai,arai} in cuprate
superconductors, these ARPES experimental results
\cite{matsui,campuzano} show that the SC coherence of the
quasiparticle peak is described by the simple
Bardeen-Cooper-Schrieffer (BCS) formalism \cite{bcs}. It is thus
established that the electron spectral function around the
$[\pi,0]$ point dramatically changes with the doping
concentration, and has a close relation to superconductivity.

Recently, we have developed a kinetic energy driven SC mechanism
\cite{feng1} based on the charge-spin separation (CSS)
fermion-spin theory \cite{feng2}, where the dressed holon-spin
interaction from the kinetic energy term induces the dressed holon
pairing state by exchanging spin excitations, then the electron
Cooper pairs originating from the dressed holon pairing state are
due to the charge-spin recombination, and their condensation
reveals the SC ground-state. In particular, this SC-state is
controlled by both SC gap function and quasiparticle coherence,
then the maximal SC transition temperature occurs around the
optimal doping, and decreases in both underdoped and overdoped
regimes \cite{feng3}. Within this framework of the kinetic energy
driven superconductivity, we \cite{feng3} have calculated the
dynamical spin structure factor, and qualitatively reproduced all
main features of inelastic neutron scattering experiments on
cuprate superconductors, including the energy dependence of the
incommensurate magnetic scattering at both low and high energies
and commensurate resonance at intermediate energy
\cite{yamada,dai,arai}. It is believed that both experiments from
ARPES and inelastic neutron scattering measurements produce
interesting data that introduce important constraints on the
microscopic models and SC theories for cuprate superconductors
\cite{shen,eschrig,fink,yamada,dai,arai}. In this Letter, we study
the electronic structure of cuprate superconductors under the
kinetic energy driven SC mechanism. Within the $t$-$t'$-$J$ model,
we have performed a systematic calculation for the electron
spectral function in the SC-state, and results show that the
spectral weight in the $[\pi,0]$ point increases with increasing
doping, and decreases with increasing temperatures. Moreover, the
position of the sharp SC quasiparticle peak in the $[\pi,0]$ point
moves to the Fermi energy as doping is increased. In analogy to
the normal-state case \cite{kim,dessau}, the SC quasiparticles
around the $[\pi,0]$ point disperse very weakly with momentum. Our
results also show that the striking behavior of the SC coherence
of the quasiparticle peaks is intriguingly related to the strong
coupling between the SC quasiparticles and collective magnetic
excitations.

In cuprate superconductors, the characteristic feature is the
presence of the CuO$_{2}$ plane \cite{bedell,kastner} as mentioned
above. It has been shown from ARPES experiments that the essential
physics of the doped CuO$_{2}$ plane is properly accounted by the
$t$-$t'$-$J$ model on a square lattice \cite{shen,wells},
\begin{eqnarray}
H&=&-t\sum_{i\hat{\eta}\sigma}C^{\dagger}_{i\sigma}
C_{i+\hat{\eta}\sigma}+t'\sum_{i\hat{\tau}\sigma}
C^{\dagger}_{i\sigma}C_{i+\hat{\tau}\sigma}\nonumber \\
&+&\mu\sum_{i\sigma}
C^{\dagger}_{i\sigma}C_{i\sigma}+J\sum_{i\hat{\eta}}{\bf S}_{i}
\cdot {\bf S}_{i+\hat{\eta}},
\end{eqnarray}
where $\hat{\eta}=\pm\hat{x},\pm \hat{y}$, $\hat{\tau}=\pm\hat{x}
\pm\hat{y}$, $C^{\dagger}_{i\sigma}$ ($C_{i\sigma}$) is the
electron creation (annihilation) operator, ${\bf S}_{i}=
C^{\dagger}_{i}{\vec\sigma} C_{i}/2$ is spin operator with
${\vec\sigma}=(\sigma_{x}, \sigma_{y},\sigma_{z})$ as Pauli
matrices, and $\mu$ is the chemical potential. This $t$-$t'$-$J$
model is subject to an important local constraint $\sum_{\sigma}
C^{\dagger}_{i\sigma}C_{i\sigma}\leq 1$ to avoid the double
occupancy. The strong electron correlation in the $t$-$t'$-$J$
model manifests itself by this local constraint \cite{anderson},
which can be treated properly in analytical calculations within
the CSS fermion-spin theory \cite{feng2}, where the constrained
electron operators are decoupled as $C_{i\uparrow}=
h^{\dagger}_{i\uparrow} S^{-}_{i}$ and $C_{i\downarrow}=
h^{\dagger}_{i\downarrow} S^{+}_{i}$, with the spinful fermion
operator $h_{i\sigma}= e^{-i\Phi_{i\sigma}}h_{i}$ describes the
charge degree of freedom together with some effects of spin
configuration rearrangements due to the presence of the doped hole
itself (dressed holon), while the spin operator $S_{i}$ describes
the spin degree of freedom (spin), then the electron local
constraint for the single occupancy, $\sum_{\sigma}
C^{\dagger}_{i\sigma}C_{i\sigma}=S^{+}_{i} h_{i\uparrow}
h^{\dagger}_{i\uparrow}S^{-}_{i}+ S^{-}_{i}h_{i\downarrow}
h^{\dagger}_{i\downarrow}S^{+}_{i}=h_{i} h^{\dagger}_{i}(S^{+}_{i}
S^{-}_{i}+S^{-}_{i}S^{+}_{i})=1- h^{\dagger}_{i}h_{i}\leq 1$, is
satisfied in analytical calculations. Moreover, these dressed
holon and spin are gauge invariant \cite{feng2}, and in this
sense, they are real and can be interpreted as the physical
excitations \cite{laughlin}. Although in common sense
$h_{i\sigma}$ is not a real spinful fermion, it behaves like a
spinful fermion. In this CSS fermion-spin representation, the
low-energy behavior of the $t$-$t'$-$J$ model (1) can be expressed
as,
\begin{eqnarray}
H&=&-t\sum_{i\hat{\eta}}(h_{i\uparrow}S^{+}_{i}
h^{\dagger}_{i+\hat{\eta}\uparrow}S^{-}_{i+\hat{\eta}}+
h_{i\downarrow}S^{-}_{i}h^{\dagger}_{i+\hat{\eta}\downarrow}
S^{+}_{i+\hat{\eta}})\nonumber\\
&+&t'\sum_{i\hat{\tau}}(h_{i\uparrow}S^{+}_{i}
h^{\dagger}_{i+\hat{\tau}\uparrow}S^{-}_{i+\hat{\tau}}+
h_{i\downarrow}S^{-}_{i}h^{\dagger}_{i+\hat{\tau}\downarrow}
S^{+}_{i+\hat{\tau}}) \nonumber \\
&-&\mu\sum_{i\sigma}h^{\dagger}_{i\sigma} h_{i\sigma}+J_{{\rm
eff}}\sum_{i\hat{\eta}}{\bf S}_{i}\cdot {\bf S}_{i+\hat{\eta}},
\end{eqnarray}
with $J_{{\rm eff}}=(1-\delta)^{2}J$, and $\delta=\langle
h^{\dagger}_{i\sigma}h_{i\sigma}\rangle=\langle h^{\dagger}_{i}
h_{i}\rangle$ is the doping concentration. As an important
consequence, the kinetic energy terms in the $t$-$t'$-$J$ model
have been transferred as the dressed holon-spin interactions, this
reflects that even the kinetic energy terms in the $t$-$t'$-$J$
Hamiltonian have strong Coulombic contributions due to the
restriction of no doubly occupancy of a given site. In cuprate
superconductors, the SC-state still is characterized by electron
Cooper pairs as in the conventional superconductors, forming SC
quasiparticles \cite{tsuei}. On the other hand, the range of the
SC gap function and pairing force in the real space have been
studied experimentally \cite{shen1,mesot}. The early ARPES
measurements \cite{shen1} showed that in the real space the gap
function and pairing force have a range of one lattice spacing.
However, the recent ARPES measurements \cite{mesot} indicated that
in the underdoped regime the presence of the higher harmonic term
${\rm cos}(2k_{x})-{\rm cos}(2k_{y})$ in the SC gap function.
Since the higher harmonic term ${\rm cos}(2k_{x})-{\rm cos}
(2k_{y})$ is closely related to the next nearest neighbor
interaction, just as the simple ${\rm cos}k_{x}-{\rm cos}k_{y}$
form in the SC gap function is closely related to the nearest
neighbor interaction, then the higher harmonics imply an increase
in the range of the pairing interaction \cite{mesot}. In other
words, the pairing interaction becomes more long range in the
underdoped regime. These higher harmonics are doping dependent,
and vanish in the overdoped regime. In particular, the SC gap
anisotropy due to the deviations from the simple ${\rm
cos}k_{x}-{\rm cos}k_{y}$ form renormalizes the slope of the
superfluid density \cite{mesot}. Although the quantitative
description of the SC properties of cuprate superconductors in the
underdoped regime needs to consider these higher harmonic effects,
the qualitative SC properties are dominated by the gap function
with the simple ${\rm cos}k_{x}-{\rm cos}k_{y}$ form \cite{mesot}.
As a qualitative discussions of the electronic structure of
cuprate superconductors in this Letter, we do not take into
account the higher harmonics in the SC gap function, and only
focus on the SC gap function with the simple ${\rm cos}k_{x}-{\rm
cos}k_{y}$ form. In this case, we can express the SC order
parameter for the electron Cooper pair in the CSS fermion-spin
representation as,
\begin{eqnarray}
\Delta &=&\langle C^{\dagger}_{i\uparrow}
C^{\dagger}_{i+\hat{\eta}\downarrow}- C^{\dagger}_{i\downarrow}
C^{\dagger}_{i+\hat{\eta}\uparrow}\rangle\nonumber\\
&=&\langle h_{i\uparrow} h_{i+\hat{\eta}\downarrow}S^{+}_{i}
S^{-}_{i+\hat{\eta}}- h_{i\downarrow}h_{i+\hat{\eta}\uparrow}
S^{-}_{i} S^{+}_{i+\hat{\eta}}\rangle\nonumber\\
&=&-\chi_{1}\Delta_{h},
\end{eqnarray}
where the spin correlation function $\chi_{1}=\langle S_{i}^{+}
S_{i+\hat{\eta}}^{-}\rangle$, and dressed holon pairing order
parameter $\Delta_{h}=\langle h_{i+\hat{\eta}\downarrow}
h_{i\uparrow}-h_{i+\hat{\eta}\uparrow} h_{i\downarrow}\rangle$.
The above result in Eq. (3) shows that the SC order parameter is
determined by the dressed holon pairing amplitude, and is
proportional to the number of doped holes, and not to the number
of electrons. In this case, although the SC order parameter
measures the strength of the binding of electrons into electron
Cooper pairs, it depends on the doping concentration, and is
similar to the doping dependent behavior of the upper critical
field \cite{wen}. In superconductors, the upper critical field is
defined as the critical field that destroys the SC-state at the
zero temperature for a given doping concentration. This indicates
that the upper critical field also measures the strength of the
binding of electrons into Cooper pairs like the SC gap parameter
\cite{wen}. In other words, both SC gap parameter and upper
critical field should have a similar doping dependence \cite{wen}.
Within the Eliashberg's strong coupling theory \cite{eliashberg},
we \cite{feng1} have shown that the dressed holon-spin interaction
can induce the dressed holon pairing state (then the electron
Cooper pairing state) by exchanging spin excitations in the higher
power of the doping concentration. Following our previous
discussions \cite{feng1}, the self-consistent equations that
satisfied by the full dressed holon diagonal and off-diagonal
Green's functions are expressed as \cite{eliashberg},
\begin{mathletters}
\begin{eqnarray}
g({\bf k},\omega)&=&g^{(0)}({\bf k},\omega)+g^{(0)}({\bf k},
\omega)[\Sigma^{(h)}_{1}({\bf k},\omega)g({\bf k},\omega)
\nonumber\\
&-& \Sigma^{(h)}_{2}(-{\bf k},-\omega)\Im^{\dagger}({\bf k},
\omega)], \\
\Im^{\dagger}({\bf k},\omega)&=&g^{(0)}(-{\bf k},-\omega)
[\Sigma^{(h)}_{1}(-{\bf k},-\omega)\Im^{\dagger}(-{\bf k},-\omega)
\nonumber\\
&+&\Sigma^{(h)}_{2}(-{\bf k},-\omega)g({\bf k},\omega)],
\end{eqnarray}
\end{mathletters}
respectively, where the mean-field (MF) dressed holon diagonal
Green's function \cite{feng2} $g^{(0)-1}({\bf k},\omega)=\omega-
\xi_{{\bf k}}$, with the MF dressed holon excitation spectrum
$\xi_{{\bf k}}=Zt\chi_{1}\gamma_{{\bf k}}-Zt'\chi_{2}\gamma_{{\bf
k}}'-\mu$, where $\gamma_{{\bf k}}=(1/Z)\sum_{\hat{\eta}}e^{i{\bf
k}\cdot \hat{\eta}}$, $\gamma_{{\bf k}}'=(1/Z)\sum_{\hat{\tau}}
e^{i{\bf k} \cdot\hat{\tau}}$, $Z$ is the number of the nearest
neighbor or next nearest neighbor sites, the spin correlation
function $\chi_{2}=\langle S_{i}^{+} S_{i+\hat{\tau}}^{-}\rangle$,
while the dressed holon self-energies are obtained from the spin
bubble as \cite{feng1,feng3},
\begin{mathletters}
\begin{eqnarray}
\Sigma^{(h)}_{1}(&{\bf k}&,i\omega_{n}) = {1\over N^{2}}
\sum_{{\bf p,p'}}\Lambda^{2}_{{\bf p+p'+k}}{1\over \beta}
\sum_{ip_{m}}g({\bf p+k},ip_{m}+i\omega_{n}) \nonumber \\
&\times&{1\over\beta}\sum_{ip'_{m}}D^{(0)}({\bf p'},ip'_{m})
D^{(0)}({\bf p'+p},ip'_{m}+ip_{m}), \\
\Sigma^{(h)}_{2}(&{\bf k}&,i\omega_{n}) = {1\over N^{2}}
\sum_{{\bf p,p'}}\Lambda^{2}_{{\bf p+p'+k}}{1\over \beta}
\sum_{ip_{m}}\Im({\bf p+k},ip_{m}+i\omega_{n})\nonumber \\
&\times&{1\over\beta}\sum_{ip'_{m}}D^{(0)}({\bf p'},ip'_{m})
D^{(0)}({\bf p'+p},ip'_{m}+ip_{m}),
\end{eqnarray}
\end{mathletters}
with $\Lambda_{{\bf k}}=Zt\gamma_{{\bf k}}-Zt'\gamma_{{\bf k}}'$,
$N$ is the number of sites, and the MF spin Green's function
\cite{feng1,feng3},
\begin{eqnarray}
D^{(0)}({\bf p},\omega)={B_{{\bf p}}\over 2\omega_{{\bf p}}} \left
({1\over\omega-\omega_{{\bf p}}}-{1\over\omega+\omega_{{\bf p}}}
\right ),
\end{eqnarray}
where $B_{{\bf p}}=2 \lambda_{1}(A_{1} \gamma_{{\bf p}}-A_{2})-
\lambda_{2} (2\chi^{z}_{2}\gamma_{{\bf p }}'-\chi_{2})$,
$\lambda_{1}= 2ZJ_{eff}$, $\lambda_{2}=4Z\phi_{2} t'$,  $A_{1}=
\epsilon \chi^{z}_{1}+\chi_{1}/2$, $A_{2} =\chi^{z}_{1}+\epsilon
\chi_{1}/2$, $\epsilon=1+2t\phi_{1} /J_{{\rm eff}}$, the dressed
holon's particle-hole parameters $\phi_{1}=\langle
h^{\dagger}_{i\sigma}h_{i+\hat{\eta}\sigma}\rangle$ and $\phi_{2}=
\langle h^{\dagger}_{i\sigma} h_{i+\hat{\tau}\sigma}\rangle$, the
spin correlation functions $\chi^{z}_{1}=\langle S_{i}^{z}
S_{i+\hat{\eta}}^{z}\rangle$ and $\chi^{z}_{2}=\langle S_{i}^{z}
S_{i+\hat{\tau}}^{z}\rangle$, and the MF spin excitation spectrum,
\begin{eqnarray}
\omega^{2}_{{\bf p}}&=&\lambda_{1}^{2}[(A_{4}-\alpha\epsilon
\chi^{z}_{1}\gamma_{{\bf p}}-{1\over 2Z}\alpha\epsilon\chi_{1})
(1-\epsilon\gamma_{{\bf p}})\nonumber \\
&+&{1\over 2}\epsilon(A_{3}-{1\over 2} \alpha\chi^{z}_{1}-\alpha
\chi_{1}\gamma_{{\bf p}})(\epsilon-
\gamma_{{\bf p}})] \nonumber \\
&+&\lambda_{2}^{2}[\alpha(\chi^{z}_{2}\gamma_{{\bf p}}'-{3\over
2Z}\chi_{2})\gamma_{{\bf p}}'+{1\over 2}(A_{5}-{1\over 2}\alpha
\chi^{z}_{2})]\nonumber\\
&+&\lambda_{1}\lambda_{2}[\alpha\chi^{z}_{1}(1-\epsilon
\gamma_{{\bf p}})\gamma_{{\bf p}}'+{1\over 2}\alpha(\chi_{1}
\gamma_{{\bf p}}'-C_{3})(\epsilon- \gamma_{{\bf p}})\nonumber\\
&+&\alpha \gamma_{{\bf p}}'(C^{z}_{3}-\epsilon \chi^{z}_{2}
\gamma_{{\bf p}})-{1\over 2}\alpha\epsilon(C_{3}- \chi_{2}
\gamma_{{\bf p}})],
\end{eqnarray}
with $A_{3}=\alpha C_{1}+(1-\alpha)/(2Z)$, $A_{4}=\alpha C^{z}_{1}
+(1-\alpha)/(4Z)$, $A_{5}=\alpha C_{2}+(1-\alpha)/(2Z)$, and the
spin correlation functions
$C_{1}=(1/Z^{2})\sum_{\hat{\eta},\hat{\eta'}}\langle
S_{i+\hat{\eta}}^{+}S_{i+\hat{\eta'}}^{-}\rangle$,
$C^{z}_{1}=(1/Z^{2})\sum_{\hat{\eta},\hat{\eta'}}\langle
S_{i+\hat{\eta}}^{z}S_{i+\hat{\eta'}}^{z}\rangle$,
$C_{2}=(1/Z^{2})\sum_{\hat{\tau},\hat{\tau'}}\langle
S_{i+\hat{\tau}}^{+}S_{i+\hat{\tau'}}^{-}\rangle$,
$C_{3}=(1/Z)\sum_{\hat{\tau}}\langle S_{i+\hat{\eta}}^{+}
S_{i+\hat{\tau}}^{-}\rangle$, and $C^{z}_{3}=(1/Z)
\sum_{\hat{\tau}}\langle S_{i+\hat{\eta}}^{z}
S_{i+\hat{\tau}}^{z}\rangle$. In order to satisfy the sum rule of
the correlation function $\langle S^{+}_{i}S^{-}_{i}\rangle=1/2$
in the case without AFLRO, an important decoupling parameter
$\alpha$ has been introduced in the MF calculation
\cite{feng1,feng2}, which can be regarded as the vertex
correction.

In the previous discussions \cite{feng1,feng3}, we have shown that
both doping and temperature dependence of the pairing force and
dressed holon gap function are incorporated into the self-energy
function $\Sigma^{(h)}_{2}({\bf k},\omega)$. In this case, the
self-energy function $\Sigma^{(h)}_{2}({\bf k},\omega)$ describes
the effective dressed holon pair gap function. On the other hand,
the self-energy function $\Sigma^{(h)}_{1}({\bf k},\omega)$
renormalizes the MF dressed holon spectrum, and thus it describes
the quasiparticle coherence. Furthermore, $\Sigma^{(h)}_{2}({\bf
k},\omega)$ is an even function of $\omega$, while
$\Sigma^{(h)}_{1}({\bf k}, \omega)$ is not. For the convenience,
we break $\Sigma^{(h)}_{1}({\bf k},\omega)$ up into its symmetric
and antisymmetric parts as, $\Sigma^{(h)}_{1} ({\bf k},\omega)=
\Sigma^{(h)}_{1e}({\bf k},\omega)+\omega \Sigma^{(h)}_{1o}({\bf
k},\omega)$, then both $\Sigma^{(h)}_{1e} ({\bf k},\omega)$ and
$\Sigma^{(h)}_{1o}({\bf k},\omega)$ are even functions of
$\omega$. According to the Eliashberg's strong coupling theory
\cite{eliashberg}, we define the quasiparticle coherent weight as
$Z^{-1}_{F}({\bf k},\omega)=1- \Sigma^{(h)}_{1o}({\bf k},\omega)$.
Since we only discuss the low-energy behavior of cuprate
superconductors, then the effective dressed holon pair gap
function and quasiparticle coherent weight can be discussed in the
static limit, i.e., $\bar{\Delta}_{h}({\bf k})=\Sigma^{(h)}_{2}
({\bf k},\omega) \mid_{\omega=0}$, $Z^{-1}_{F}({\bf k})=1-
\Sigma^{(h)}_{1o}({\bf k},\omega)\mid_{\omega=0}$, and
$\Sigma^{(h)}_{1e} ({\bf k})= \Sigma^{(h)}_{1e}({\bf k},\omega)
\mid_{\omega=0}$. Although $Z_{F}({\bf k})$ and
$\Sigma^{(h)}_{1e}({\bf k})$ still are a function of ${\bf k}$,
the wave vector dependence may be unimportant \cite{eliashberg}.
From ARPES experiments \cite{shen,eschrig,fink}, it has been shown
that in the SC-state, the lowest energy states are located at the
$[\pi,0]$ point, which indicates that the majority contribution
for the electron spectrum comes from the $[\pi,0]$ point. In this
case, the wave vector ${\bf k}$ in $Z_{F}({\bf k})$ and
$\Sigma^{(h)}_{1e}({\bf k})$ can be chosen as $Z^{-1}_{F}=
1-\Sigma^{(h)}_{1o}({\bf k})\mid_{{\bf k}=[\pi,0]}$ and
$\Sigma^{(h)}_{1e}=\Sigma^{(h)}_{1e}({\bf k})\mid_{{\bf k}=
[\pi,0]}$. This is different from the previous discussions
\cite{feng3}, where the wave vector ${\bf k}$ in $Z_{F}({\bf k})$
and $\Sigma^{(h)}_{1e}({\bf k})$ has been chosen near the nodal
point in the Brillouin zone. With the help of the above
discussions, the dressed holon diagonal and off-diagonal Green's
functions in Eqs. (4a) and (4b) can be obtained explicitly as,
\begin{mathletters}
\begin{eqnarray}
g({\bf k},\omega)&=&Z_{F}{U^{2}_{h{\bf k}}\over\omega-E_{h{\bf k}}
}+Z_{F}{V^{2}_{h{\bf k}}\over\omega+E_{h{\bf k}}}, \\
\Im^{\dagger}({\bf k},\omega)&=&-Z_{F}{\bar{\Delta}_{hZ}({\bf k})
\over 2E_{h{\bf k}}}\left ( {1\over\omega-E_{h{\bf k}}}-{1\over
\omega + E_{h{\bf k}}}\right ),
\end{eqnarray}
\end{mathletters}
where the dressed holon quasiparticle coherence factors
$U^{2}_{h{\bf k}}=(1+\bar{\xi_{{\bf k}}}/E_{h{\bf k}})/2$ and
$V^{2}_{h{\bf k}}=(1-\bar{\xi_{{\bf k}}}/E_{h{\bf k}})/2$, the
renormalized dressed holon excitation spectrum $\bar{\xi_{{\bf k}
}}=Z_{F}(\xi_{{\bf k}}+\Sigma^{(h)}_{1e})$, the renormalized
dressed holon pair gap function $\bar{\Delta}_{hZ}({\bf k})=Z_{F}
\bar{\Delta}_{h}({\bf k})$, and the dressed holon quasiparticle
spectrum $E_{h{\bf k}}= \sqrt {\bar{\xi^{2}_{{\bf k}}}+\mid
\bar{\Delta}_{hZ}({\bf k})\mid^{2}}$. This dressed holon
quasiparticle is the excitation of a single dressed holon
"adorned" with the attractive interaction between paired dressed
holons, while the $Z_{F}$ reduces the dressed holon (then
electron) quasiparticle bandwidth, and then the energy scale of
the electron quasiparticle band is controlled by the magnetic
interaction $J$. On the other hand, cuprate superconductors are
characterized by an overall d-wave pairing symmetry \cite{tsuei}.
In particular, we \cite{feng3} have shown within the $t$-$J$ type
model that the electron Cooper pairs have a dominated d-wave
symmetry over a wide range of the doping concentration, around the
optimal doping. Therefore in the following discussions, we
consider the d-wave case, i.e., $\bar{\Delta}_{hZ}({\bf k})=
\bar{\Delta}_{hZ}\gamma^{(d)}_{{\bf k}}$, with $\gamma^{(d)}_{{\bf
k}}=({\rm cos} k_{x}-{\rm cos}k_{y})/2$. In this case, the dressed
holon effective gap parameter and quasiparticle coherent weight in
Eqs. (5a) and (5b) satisfy the following two equations,
\begin{mathletters}
\begin{eqnarray}
1&=&{1\over N^{3}}\sum_{{\bf k,p,p'}}\Lambda^{2}_{{\bf p+k}}
\gamma^{(d)}_{{\bf k-p'+p}}\gamma^{(d)}_{{\bf k}}{Z^{2}_{F}\over
E_{h{\bf k}}}{B_{{\bf p}}B_{{\bf p'}}\over\omega_{{\bf p}}
\omega_{{\bf p'}}}\nonumber \\
&\times&\left({F^{(1)}_{1}({\bf k,p,p'})\over (\omega_{{\bf p'}}
-\omega_{{\bf p}})^{2}-E^{2}_{h{\bf k}}}- {F^{(2)}_{1}({\bf k,p,
p'})\over (\omega_{{\bf p'}}+\omega_{{\bf p}})^{2}-
E^{2}_{h{\bf k}}} \right ) ,\\
{1\over Z}_{F} &=& 1+ {1\over N^{2}}\sum_{{\bf p,p'}}
\Lambda^{2}_{{\bf p}+{\bf k}_{0}} Z_{F} {B_{{\bf p}}B_{{\bf p'}}
\over 4\omega_{{\bf p}} \omega_{{\bf p'}}}\nonumber\\
&\times&\left ( {F^{(1)}_{2}({\bf p,p'})\over (\omega_{{\bf p}}-
\omega_{{\bf p'}}- E_{h{\bf p-p'+k_{0}}} )^{2}}\right .
\nonumber\\
&+&{F^{(2)}_{2}({\bf p,p'})\over (\omega_{{\bf p}}-\omega_{{\bf
p'}}+E_{h{\bf p-p'+k_{0}}})^{2}}\nonumber \\
&+&{F^{(3)}_{2}({\bf p,p'}) \over (\omega_{{\bf p}}+ \omega_{{\bf
p'}}- E_{h{\bf p-p'+k_{0}}})^{2}}\nonumber \\
&+&\left . {F^{(4)}_{2} ({\bf p, p'}) \over (\omega_{{\bf p}} +
\omega_{{\bf p'}} + E_{h{\bf p-p'+k_{0} }})^{2}} \right ) ,
\end{eqnarray}
\end{mathletters}
respectively, where ${\bf k}_{0}=[\pi,0]$, $F^{(1)}_{1}({\bf k,p,
p'})=(\omega_{{\bf p'}}- \omega_{{\bf p}})[n_{B}(\omega_{{\bf p}})
-n_{B}(\omega_{{\bf p'}}) ][1-2 n_{F}(E_{h{\bf k}})]+E_{h{\bf k}}
[n_{B}(\omega_{{\bf p'}}) n_{B}(-\omega_{{\bf p}})+ n_{B}
(\omega_{{\bf p}}) n_{B} (-\omega_{{\bf p'}})]$, $F^{(2)}_{1}
({\bf k,p,p'})=(\omega_{{\bf p'}}+\omega_{{\bf p}})[n_{B}(-
\omega_{{\bf p'}}) - n_{B} (\omega_{{\bf p}})] [1-2 n_{F}
(E_{h{\bf k}})]+E_{h{\bf k}}[n_{B} (\omega_{{\bf p'}})n_{B}
(\omega_{{\bf p}}) + n_{B}(-\omega_{{\bf p'}}) n_{B}
(-\omega_{{\bf p}})]$,   $F^{(1)}_{2}({\bf p,p'}) = n_{F}
(E_{h{\bf p-p'+k_{0}}})[n_{B}(\omega_{{\bf p'}})- n_{B}
(\omega_{{\bf p}})] - n_{B}(\omega_{{\bf p}})  n_{B}
(-\omega_{{\bf p'}})$,  $F^{(2)}_{2} ({\bf p,p'}) = n_{F}
(E_{h{\bf p-p'+k_{0}}}) [n_{B}(\omega_{{\bf p}})-n_{B}
(\omega_{{\bf p'}})] - n_{B} (\omega_{{\bf p'}})n_{B}(-
\omega_{{\bf p}})$, $F^{(3)}_{2}({\bf p,p'})= n_{F} (E_{h{\bf
p-p'+k_{0}}})[n_{B}(\omega_{{\bf p'}})- n_{B}(-\omega_{{\bf
p}})]+n_{B}(\omega_{{\bf p}})n_{B} (\omega_{{\bf p'}})$, and
$F^{(4)}_{2}({\bf p,p'})=n_{F}(E_{h{\bf p-p'+k_{0}}})[n_{B}
(-\omega_{{\bf p'}})-n_{B}(\omega_{{\bf p}})]+ n_{B}(-\omega_{{\bf
p}})n_{B}(-\omega_{{\bf p'}})$. These two equations must be solved
simultaneously with other self-consistent equations
\cite{feng1,feng2,feng3}, then all order parameters, decoupling
parameter $\alpha$, and chemical potential $\mu$ are determined by
the self-consistent calculation. In this sense, our above
calculations are controllable without using adjustable parameters.

According to the dressed holon off-diagonal Green's function (8b),
the dressed holon pair gap function is obtained as $\Delta_{h}
({\bf k})=-(1/\beta)\sum_{i\omega_{n}}\Im^{\dagger}({\bf k},
i\omega_{n})$. In the previous discussions \cite{feng1}, it has
been shown that this dressed holon pairing state originating from
the kinetic energy terms by exchanging spin excitations also leads
to form the electron Cooper pairing state. For discussions of the
electronic structure in the SC-state, we need to calculate the
electron diagonal and off-diagonal Green's functions $G(i-j,t-t')
=\langle\langle C_{i\sigma}(t); C^{\dagger}_{j\sigma} (t')\rangle
\rangle$ and $\Gamma^{\dagger} (i-j,t-t')=\langle \langle
C^{\dagger}_{i\uparrow}(t); C^{\dagger}_{j\downarrow}(t')\rangle
\rangle$, which are the convolutions of the spin Green's function
and dressed holon diagonal and off-diagonal Green's functions in
the CSS fermion-spin theory, and can be obtained in terms of the
MF spin Green's function (6) and dressed holon diagonal and
off-diagonal Green's functions (8a) and (8b) as,
\begin{mathletters}
\begin{eqnarray}
G({\bf k},\omega)&=&{1\over N}\sum_{{\bf p}}Z_{F}{B_{{\bf p}}\over
4\omega_{{\bf p}}}\left \{ {\rm coth}[{1\over 2}\beta\omega_{{\bf
p}}]\right .\nonumber \\
&\times&\left ( {U^{2}_{h{\bf p+k}}\over\omega+E_{h{\bf p+k}}-
\omega_{{\bf p}}}+{U^{2}_{h{\bf p+k}}\over\omega+E_{h{\bf
p+k}}+\omega_{{\bf p}}} \right .\nonumber \\
&+& \left . {V^{2}_{h{\bf p+k}}\over\omega-E_{h{\bf p+k}}+
\omega_{{\bf p}}}+{V^{2}_{h{\bf p+k}}\over\omega-E_{h{\bf p+k}}
-\omega_{{\bf p}}} \right )\nonumber \\
&+&{\rm tanh}[{1\over 2}\beta E_{h{\bf p+k}}]\left ({U^{2}_{h{\bf
p+k}}\over\omega+E_{h{\bf p+k}}+\omega_{{\bf p}}}\right .
\nonumber \\
&-&{U^{2}_{h{\bf p+k}}\over\omega+E_{h{\bf p+k}}-\omega_{{\bf p}}}
+{V^{2}_{h{\bf p+k}}\over\omega-E_{h{\bf p+k}}- \omega_{{\bf p}}}
\nonumber \\
&-&\left .\left .{V^{2}_{h{\bf p+k}}\over\omega-E_{h{\bf p+k}}
+\omega_{{\bf p}}} \right ) \right \} ,\\
\Gamma^{\dagger}({\bf k},\omega)&=&{1\over N}\sum_{{\bf p}}
Z_{F}{\bar{\Delta}_{hZ}({\bf p+k})\over 2E_{h{\bf p+k}}}{B_{{\bf p
}}\over 4\omega_{{\bf p}}}\left \{{\rm coth}[{1\over
2}\beta\omega_{{\bf p}}] \right . \nonumber \\
&\times&\left ({1\over \omega-E_{h{\bf p+k}} -\omega_{{\bf p}}}
+{1\over \omega-E_{h{\bf p+k}}+\omega_{{\bf p}}}\right.\nonumber\\
&-&\left . {1\over\omega +E_{h{\bf p+k}}+\omega_{{\bf p}}}-
{1\over \omega +E_{h{\bf p+k}}-\omega_{{\bf p}}} \right
)\nonumber\\
&+&{\rm tanh}[{1\over 2}\beta E_{h{\bf p+k}}]\left ({1\over \omega
-E_{h{\bf p+k}} -\omega_{{\bf p}}} \right .\nonumber \\
&-&{1\over \omega-E_{h{\bf p+k}}+\omega_{{\bf p}}} -{1\over \omega
+E_{h{\bf p+k}}+\omega_{{\bf p}}}\nonumber\\
&+& \left . \left .{1\over\omega +E_{h{\bf p+k}}-\omega_{{\bf p}}}
\right )\right \},
\end{eqnarray}
\end{mathletters}
respectively, these convolutions of the spin Green's function and
dressed holon diagonal and off-diagonal Green's functions reflect
the charge-spin recombination \cite{anderson1}, then the electron
spectral function $A({\bf k},\omega)=-2{\rm Im}G({\bf k},\omega)$
and SC gap function $\Delta({\bf k})=-(1/\beta)\sum_{i\omega_{n}}
\Gamma^{\dagger}({\bf k},i\omega_{n})$ are obtained from the above
electron diagonal and off-diagonal Green's functions as,
\begin{mathletters}
\begin{eqnarray}
A({\bf k},\omega)&=&2\pi {1\over N}\sum_{{\bf p}}Z_{F}{B_{{\bf p}}
\over 4\omega_{{\bf p}}}\left \{{\rm coth}({1\over 2}\beta
\omega_{{\bf p}})\right . \nonumber\\
&\times& [U^{2}_{h{\bf p+k}}\delta(\omega+E_{h{\bf p+k}}
-\omega_{{\bf p }})\nonumber \\
&+&U^{2}_{h{\bf p+k}}\delta(\omega+E_{h{\bf p+k}}+\omega_{{\bf p}}
)\nonumber\\
&+&V^{2}_{h{\bf p+k}}\delta(\omega-E_{h{\bf p+k}}+\omega_{{\bf p}}
)\nonumber \\
&+&V^{2}_{h{\bf p+k}}\delta(\omega-E_{h{\bf p+k}}-\omega_{{\bf p}}
)] \nonumber \\
&+&{\rm tanh}({1\over 2}\beta E_{h{\bf p+k}})[U^{2}_{h{\bf p+k}}
\delta(\omega+E_{h{\bf p+k}}+\omega_{{\bf p}})\nonumber \\
&-&U^{2}_{h{\bf p+k}}\delta(\omega+E_{h{\bf p+k}}-\omega_{{\bf
p}})\nonumber \\
&+& V^{2}_{h{\bf p+k}}\delta(\omega-E_{h{\bf p+k}}-
\omega_{{\bf p}})\nonumber\\
&-&\left . V^{2}_{h{\bf p+k}}\delta(\omega-E_{h{\bf p+k}}
+\omega_{{\bf p}})]\right \} ,\\
\Delta({\bf k})&=&-{1\over N}\sum_{{\bf p}}{Z_{F}
\bar{\Delta}_{Zh}({\bf p-k})\over 2E_{h{\bf p-k}}}{\rm tanh}
[{1\over 2}\beta E_{h{\bf p-k}}]\nonumber \\
&\times&{B_{{\bf p}}\over 2\omega_{{\bf p}}}{\rm coth} [{1\over
2}\beta\omega_{{\bf p}}],
\end{eqnarray}
\end{mathletters}
respectively. With the above SC gap function (11b), the SC gap
parameter in Eq. (3) is obtained as $\Delta=-\chi_{1}\Delta_{h}$.
Since both dressed holon (then electron) pairing gap parameter and
pairing interaction in cuprate superconductors are doping
dependent, then the experimental observed doping dependence of the
SC gap parameter should be an effective SC gap parameter
$\bar{\Delta}\sim-\chi_{1}\bar{\Delta}_{h}$. For a complement of
the previous analysis of the interplay between the quasiparticle
coherence and superconductivity \cite{feng3}, we plot (a) the
quasiparticle coherent weight $Z_{F}(T_{c})$, (b) the effective SC
gap parameter $\bar{\Delta}$ at temperature $T=0.002J$, and (c)
the SC transition temperature $T_{c}$ as a function of the doping
concentration for parameters $t/J=2.5$ and $t'/J=0.3$ in Fig. 1.
For comparison, the corresponding experimental results of the
quasiparticle coherent weight in the $[\pi,0]$ point \cite{ding},
upper critical field \cite{wen}, and SC transition temperature
\cite{tallon} as a function of the doping concentration are also
shown in Fig. 1(a), 1(b), and 1(c), respectively. Although we
focus on the quasiparticle coherent weight at the antinodal point
in the above discussions, our present results of the doping
dependence of the effective SC gap parameter and SC transition
temperature are consistent with these of the previous results
\cite{feng3}, where it has been focused on the quasiparticle
coherent weight near the nodal point. The quasiparticle coherent
weight grows linearly with the doping concentration, i.e.,
$Z_{F}\propto\delta$, which together with the SC gap parameter
defined in Eq. (3) show that only $\delta$ number of coherent
doped carriers are recovered in the SC-state, consistent with the
picture of a doped Mott insulator with $\delta$ holes
\cite{anderson}. In this case, the SC-state of cuprate
superconductors is controlled by both SC gap function and
quasiparticle coherence \cite{ding,feng3}, then the SC transition
temperature increases with increasing doping in the underdoped
regime, and reaches a maximum in the optimal doping, then
decreases in the overdoped regime. Using an reasonably estimative
value of $J\sim 800$K to 1200K in cuprate superconductors
\cite{bedell,kastner}, the SC transition temperature in the
optimal doping is T$_{c}\approx 0.165J \approx 132{\rm K}\sim
198{\rm K}$, in qualitative agreement with the experimental data
\cite{tallon}.

\begin{figure}[prb]
\epsfxsize=3.5in\centerline{\epsffile{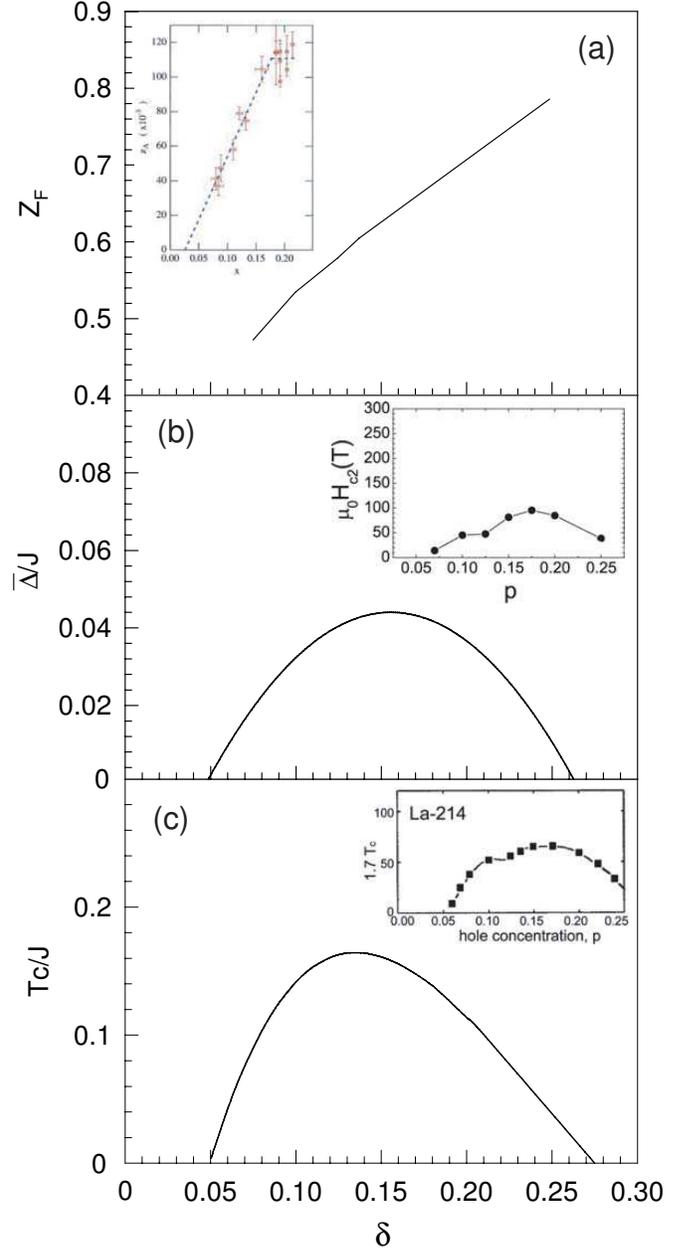}}\caption{(a) The
quasiparticle coherent weight $Z_{F}(T_{c})$ in the $[\pi,0]$
point, (b) the effective SC gap parameter $\bar{\Delta}$ at
$T=0.002J$, and (c) the SC transition temperature $T_{c}$ as a
function of the doping concentration for $t/J=2.5$ and $t'/J=0.3$.
Inset: the corresponding experimental results of cuprate
superconductors taken from Refs. [11], [27], and [30],
respectively.}
\end{figure}

\begin{figure}[prb]
\epsfxsize=3.5in\centerline{\epsffile{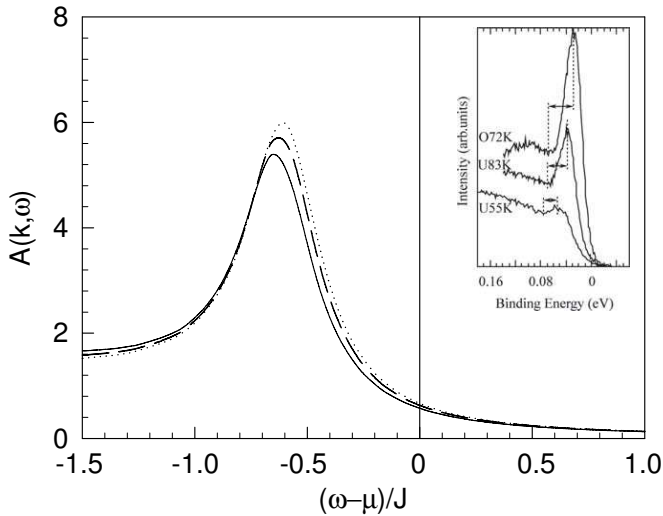}} \caption{The
electron spectral function $A({\bf k},\omega)$ in the $[\pi,0]$
point with $\delta=0.09$ (solid line), $\delta=0.12$ (dashed
line), and $\delta=0.15$ (dotted line) at $T=0.002J$ for $t/J=2.5$
and $t'/J=0.3$.  Inset: the experimental result of cuprate
superconductors taken from Ref. [9].}
\end{figure}

Now we are ready to discuss the electronic structure of cuprate
superconductors. We have performed a calculation for the electron
spectral function (11a), and the results of $A({\bf k},\omega)$ in
the $[\pi,0]$ point with the doping concentration $\delta=0.09$
(solid line), $\delta=0.12$ (dashed line), and $\delta=0.15$
(dotted line) at temperature $T=0.002J$ for parameters $t/J=2.5$
and $t'/J=0.3$ are plotted in Fig. 2 in comparison with the
experimental result \cite{campuzano1} (inset). Our results show
that there is a sharp SC quasiparticle peak near the electron
Fermi energy in the $[\pi,0]$ point, and the position of the SC
quasiparticle peak in the doping concentration $\delta=0.15$ is
located at $\omega_{{\rm peak}}\approx 0.6J\approx 0.042$eV$\sim
0.06$eV, which is qualitatively consistent with $\omega_{{\rm
peak}}\approx 0.03$eV observed \cite{campuzano1,ding,shen} in the
slightly overdoped cuprate superconductor
Bi$_{2}$Sr$_{2}$CaCu$_{2}$O$_{8+x}$. Moreover, the electron
spectrum is doping dependent. With increasing the doping
concentration, the weight of the SC quasiparticle peaks increases,
while the position of the SC quasiparticle peak moves to the Fermi
energy \cite{campuzano1,shen}. Furthermore, we have discussed the
temperature dependence of the electron spectrum, and the results
of $A({\bf k},\omega)$ in the $[\pi,0]$ point with the doping
concentration $\delta=0.15$ at temperature $T=0.002J$ (solid
line), $T=0.10J$ (dashed line), and $T=0.15J$ (dotted line) for
parameters $t/J=2.5$ and $t'/J=0.3$ are plotted in Fig. 3 in
comparison with the experimental result \cite{dlfeng} (inset).
These results show that the spectral weight decreases as
temperature is increased, in qualitative agreement with the
experimental data \cite{dlfeng,shen}. This temperature dependence
of the electron spectrum in cuprate superconductors has also been
discussed in Ref. \cite{norman}. By direct analysis of the ARPES
data, they \cite{norman} studied the temperature dependence of the
electron self-energy, and then indicated that the spectral
lineshape in the $[\pi,0]$ point is naturally explained by the
coupling of the electrons to a magnetic resonance. Since the
intensity of this resonance decreases with temperature, then the
coupling of the electrons to this magnetic mode also decreases. As
the magnetic resonance intensity decreases, the spin gap in the
dynamic susceptibility fills in, which may be responsible for the
"filling in" of the imaginary part of the electron self-energy
\cite{norman}. The combination of these two effects cause the
spectral peak to rapidly broaden with temperature. Our results are
also consistent with their results.

\begin{figure}[prb]
\epsfxsize=3.5in\centerline{\epsffile{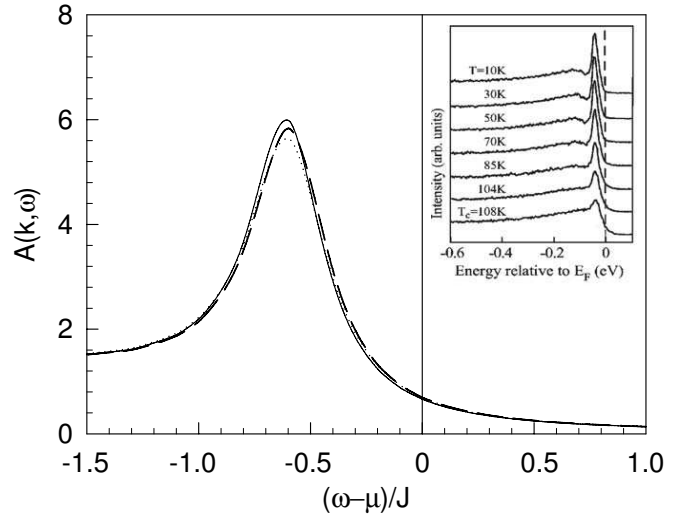}}\caption{The
electron spectral function $A({\bf k},\omega)$ in the $[\pi,0]$
point with $\delta=0.15$ at $T=0.002J$ (solid line), $T=0.10J$
(dashed line), and $T=0.15J$ (dotted line) for $t/J=2.5$ and
$t'/J=0.3$. Inset: the experimental result of cuprate
superconductors taken from Ref. [8].}
\end{figure}

\begin{figure}[prb]
\epsfxsize=3.5in\centerline{\epsffile{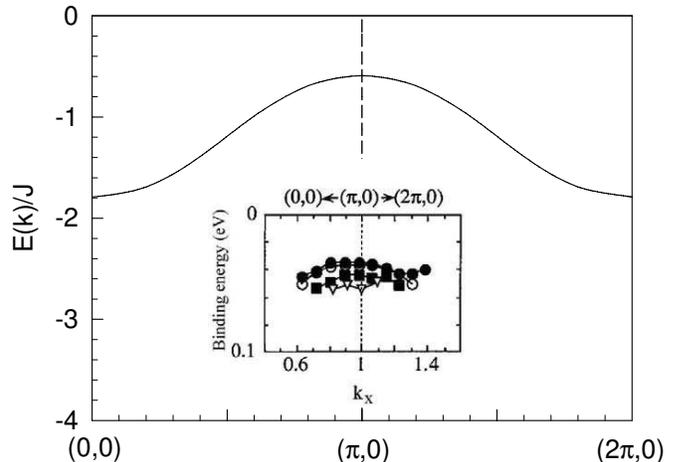}} \caption{The
positions of the lowest energy SC quasiparticle peaks in $A({\bf
k},\omega)$ as a function of momentum along the direction
$[0,0]\rightarrow [\pi,0]\rightarrow [2\pi,0]$ with $\delta=0.15$
at $T=0.002J$ for $t/J=2.5$ and $t'/J=0.3$. Inset: the
experimental result of cuprate superconductors taken from Ref.
[9].}
\end{figure}

For a better understanding of the anomalous form of the electron
spectrum $A({\bf k},\omega)$ as a function of energy $\omega$ for
${\bf k}$ in the vicinity of the $[\pi,0]$ point, we have made a
series of calculations for $A({\bf k},\omega)$ around the
$[\pi,0]$ point, and the results show that the sharp SC
quasiparticle peak persists in a very large momentum space region
around the $[\pi,0]$ point. To show this point clearly, we plot
the positions of the lowest energy SC quasiparticle peaks in
$A({\bf k},\omega)$ as a function of momentum along the direction
$[0,0] \rightarrow [\pi,0]\rightarrow [2\pi,0]$ at the doping
concentration $\delta=0.15$ with temperature $T=0.002J$ for
parameters $t/J=2.5$ and $t'/J=0.3$ in Fig. 4 in comparison with
the experimental result \cite{campuzano1} of the cuprate
superconductor Bi$_{2}$Sr$_{2}$CaCu$_{2}$O$_{8+\delta}$ (inset).
It is shown that the sharp SC quasiparticle peaks around the
$[\pi,0]$ point at low energies disperse very weakly with
momentum, which is corresponding to the unusual flat band appeared
in the normal-state around the $[\pi,0]$ point \cite{kim,dessau},
and is qualitatively consistent with these obtained from ARPES
experimental measurements on doped cuprates
\cite{shen,eschrig,fink,campuzano1}.

A nature question is why the SC coherence of the SC quasiparticle
peak in cuprate superconductors can be described qualitatively in
the framework of the kinetic energy driven superconductivity. The
reason is that the SC-state in the kinetic energy driven
superconductivity is the conventional BCS like \cite{feng3}. This
can be understood from the electron diagonal and off-diagonal
Green's functions in Eqs. (10a) and (10b). Since the spins center
around the $[\pi,\pi]$ point in the MF level \cite{feng1,feng2},
then the main contributions for the spins comes from the
$[\pi,\pi]$ point. In this case, the electron diagonal and
off-diagonal Green's functions in Eqs. (10a) and (10b) can be
approximately reduced in terms of $\omega_{{\bf p}=[\pi,\pi]}\sim
0$ and the equation \cite{feng1,feng2} $1/2=\langle S_{i}^{+}
S_{i}^{-}\rangle =(1/N)\sum_{{\bf p}}B_{{\bf p}}{\rm coth}(\beta
\omega_{{\bf p}}/2)/(2\omega_{{\bf p}})$ as,
\begin{mathletters}
\begin{eqnarray}
g({\bf k},\omega)&\approx&Z_{F}{U^{2}_{{\bf k}}\over \omega-
E_{{\bf k}}}+Z_{F}{V^{2}_{{\bf k}}\over \omega+E_{{\bf k}}},\\
\Gamma^{\dagger}({\bf k},\omega)&\approx& Z_{F}{\bar{\Delta}_{hZ}
({\bf k })\over 2E_{{\bf k}}}\left ({1\over \omega-E_{{\bf k}}}+
{1\over \omega+E_{{\bf k}}}\right ),
\end{eqnarray}
\end{mathletters}
where the electron quasiparticle coherence factors $U^{2}_{{\bf k}
}\approx V^{2}_{h{\bf k+k_{A}}}$ and $V^{2}_{{\bf k}}\approx
U^{2}_{h{\bf k+k_{A}}}$, and electron quasiparticle spectrum
$E_{{\bf k}}\approx E_{h{\bf k+k_{A}}}$, with ${\bf k_{A}}=
[\pi,\pi]$, i.e., the hole-like dressed holon quasiparticle
coherence factors $V_{h{\bf k}}$ and $U_{h{\bf k}}$ and hole-like
dressed holon quasiparticle spectrum $E_{h{\bf k}}$ have been
transferred into the electron quasiparticle coherence factors
$U_{{\bf k}}$ and $V_{{\bf k}}$ and electron quasiparticle
spectrum $E_{{\bf k}}$, respectively, by the convolutions of the
spin Green's function and dressed holon Green's functions due to
the charge-spin recombination. This means that the dressed holon
pairs condense with the d-wave symmetry in a wide range of the
doping concentration, then the electron Cooper pairs originating
from the dressed holon pairing state are due to the charge-spin
recombination, and their condensation automatically gives the
electron quasiparticle character. This electron quasiparticle is
the excitation of a single electron "dressed" with the attractive
interaction between paired electrons. This is why the basic BCS
formalism \cite{bcs} is still valid in discussions of the doping
dependence of the effective SC gap parameter and SC transition
temperature, and SC coherence of the quasiparticle peak
\cite{matsui,campuzano}, although the pairing mechanism is driven
by the kinetic energy by exchanging spin excitations, and other
exotic magnetic scattering \cite{yamada,dai,arai,feng3} is beyond
the BCS theory.

In summary, we have studied the electronic structure of cuprate
superconductors based on the kinetic energy driven
superconductivity. Our results show that the spectral weight of
the electron spectrum in the $[\pi,0]$ point decreases as
temperature is increased. With increasing the doping
concentration, this spectral weight increases, while the position
of the sharp SC quasiparticle peak moves to the Fermi energy. In
analogy to the normal-state case \cite{kim,dessau}, the SC
quasiparticles around the $[\pi,0]$ point disperse very weakly
with momentum. Our results also show that the striking behavior of
the SC coherence of the quasiparticle peak is intriguingly related
to the strong coupling between the SC quasiparticles and
collective magnetic excitations.

\acknowledgments

This work was supported by the National Natural Science Foundation
of China under Grant Nos. 10125415 and 90403005, and the Grant
from the Ministry of Science and Technology of China under Grant
No. 2006CB601002.


\begin{references}

\bibitem{bedell} See, e.g., {\it  Proceedings of Los Alamos
Symposium}, edited by K.S. Bedell, D. Coffey, D.E. Meltzer, D.
Pines, and J.R. Schrieffer (Addison-Wesley, Redwood city,
California, 1990).

\bibitem{kastner} See, e.g., M.A.Kastner, R.J. Birgeneau, G.
Shiran, and Y. Endoh, Rev. Mod. Phys. {\bf 70}, 897 (1998), and
referenes therein.

\bibitem {anderson} P.W. Anderson, in {\it Frontiers and
Borderlines in Many Particle Physics}, edited by R.A. Broglia and
J.R. Schrieffer (North-Holland, Amsterdam, 1987), p. 1; Science
{\bf 235}, 1196 (1987).

\bibitem{shen} See, e.g., A. Damascelli, Z. Hussain, and Z.-X.
Shen, Rev. Mod. Phys. {\bf 75}, 475 (2003), and referenes therein.

\bibitem{eschrig} See, e.g., J. Campuzano, M. Norman, M. Randeira,
in {\it Physics of Superconductors}, vol. II, edited by K.
Bennemann and J. Ketterson (Springer, Berlin Heidelberg New York,
2004), p. 167, and referenes therein.

\bibitem{fink} See, e.g., J. Fink, S. Borisenko, A. Kordyuk, A.
Koitzsch, J. Geck, V. Zabolotnyy, M. Knupfer, B. B\"uchner, and H.
Berger, cond-mat/0512307, and referenes therein.

\bibitem{campuzano} J. Campuzano, H. Ding, M. Norman, M. Randeira,
A. Bellman, T. Mochiku, and K. Kadowaki, Phys. Rev. B {\bf 53},
14737 (2003).

\bibitem{dlfeng} D.L. Feng, A. Damascelli, K.M. Shen, N.
Motoyama, D.H. Lu, H. Eisaki, K. Shimizu, J.-i. Shimoyama, K.
Kishio, N. Kaneko, M. Greven, G.D. Gu, X.J. Zhou, C. Kim, F.
Ronning, N.P. Armitage, and Z.-X Shen, Phys. Rev. Lett. {\bf 88},
107001 (2002).

\bibitem{campuzano1} J. Campuzano, H. Ding, M. Norman, H. Fretwell,
M. Randeira, A. Kaminski, J. Mesot, T. Takeuchi, T. Sato, T.
Yokoya, T. Takahashi, T. Mochiku, K. Kadowaki, P. Guptasarma, D.
Hinks, Z. Konstantinovic, Z. Li, and H. Raffy, Phys. Rev. Lett.
{\bf 83}, 3709 (1999).

\bibitem{dewilde} Y. DeWilde, N. Miyakawa, P. Guptasarma, M.
Iavarone, L. Ozyuzer, J. Zasadzinski, P. Romano, D. Hinks, C.
Kendziora, G. Crabrtee, and K. Gray, Phys. Rev. Lett. {\bf 80},
153 (1998).

\bibitem{ding} H. Ding, J.R. Engelbrecht, Z. Wang, J.C. Campuzano,
S.C. Wang, H.B. Yang, R. Rogan, T. Takahashi, K. Kadowaki, and
D.G. Hinks, Phys. Rev. Lett. {\bf 87}, 227001 (2001).

\bibitem{matsui} H. Matsui, T. Sato, T. Takahashi, S.C. Wang, H.B.
Yang, H. Ding, T. Fujii, T. Watanabe, and A. Matsuda, Phys. Rev.
Lett. {\bf 90}, 217002 (2003).

\bibitem{yamada} K. Yamada, C.H. Lee, K. Kurahashi, J. Wada, S.
Wakimoto, S. Ueki, H. Kimura, Y. Endoh, S. Hosoya, and G. Shirane,
Phys. Rev. B {\bf 57}, 6165 (1998).

\bibitem{dai} P. Dai, H.A. Mook, R.D. Hunt, and F. Do\~gan, Phys.
Rev. B{\bf 63}, 54525 (2001); P. Bourges, B. Keimer, S. Pailh\'es,
L.P. Regnault, Y. Sidis, and C. Ulrich, Physica C {\bf 424}, 45
(2005).

\bibitem{arai} M. Arai, T. Nishijima, Y. Endoh, T. Egami, S.
Tajima, K. Tomimoto, Y. Shiohara, M. Takahashi, A. Garret, and
S.M. Bennington, Phys. Rev. Lett. {\bf 83}, 608 (1999); S.M.
Hayden, H.A. Mook, P. Dai, T.G. Perring, and F. Do\~gan, Nature
{\bf 429}, 531 (2004); C. Stock, W.J. Buyers, R.A. Cowley, P.S.
Clegg, R. Coldea, C.D. Frost, R. Liang, D. Peets, D. Bonn, W.N.
Hardy, and R.J. Birgeneau, Phys. Rev. B{\bf 71}, 24522 (2005).

\bibitem{bcs} J.R. Schrieffer, {\it Theory of Superconductivity},
Benjamin, New York, 1964.

\bibitem{feng1} Shiping Feng, Phys. Rev. {\bf B68}, 184501
(2003).

\bibitem{feng2} Shiping Feng, Jihong Qin, and Tianxing Ma, J.
Phys. Condens. Matter {\bf 16}, 343 (2004); Shiping Feng, Tianxing
Ma, and Jihong Qin, Mod. Phys. Lett. B{\bf 17}, 361 (2003).

\bibitem{feng3} Shiping Feng, Tianxing Ma, and Huaiming Guo,
Physica C {\bf 436}, 14 (2006); Shiping Feng and Tianxing Ma,
Phys. Lett. A {\bf 350}, 138 (2006); Shiping Feng and Tianxing Ma,
in {\it New Frontiers in Superconductivity Research}, edited by
B.P. Martins (Nova Science Publishers, New York, 2006), Chapter
12, in press, cond-mat/0603148.

\bibitem{kim} Z.X. Shen, W.E. Spicer, D.M. King, D.S. Dessau, and
B.O. Wells, Science {\bf 267}, 343 (1995); D.M. King, Z.X. Shen,
D.S. Dessau, D.S. Marshall, C.H. Park, W.E. Spicer, J.L. Peng,
Z.Y. Li, and R.L. Greene, Phys. Rev. Lett. {\bf 73}, 3298 (1994).

\bibitem{dessau} D.S. Dessau, Z.X. Shen, D.M. King, D.S. Marshall,
L.W. Lombardo, P.H. Dickinson, A.G. Loeser, J. DiCarlo, C.H. Park,
A. Kapitulnik, and W.E. Spicer, Phys. Rev. Lett. {\bf 71}, 2781
(1993); Huaiming Guo and Shiping Feng, Phys. Lett. A {\bf 355},
473 (2006).

\bibitem{wells} B.O. Wells, Z.X. Shen, A. Matsuura, D.M. King,
M. A. Kastner, M. Greven, and R.J. Birgeneau, Phys. Rev. Lett.
{\bf 74}, 964 (1995); C. Kim, P.J. White, Z.X. Shen, T. Tohyama,
Y. Shibata, S. Maekawa, B.O. Wells, Y.J. Kim, R.J. Birgeneau, and
M.A. Kastner, Phys. Rev. Lett. {\bf 80}, 4245 (1998).

\bibitem{laughlin} R.B Laughlin, Phys. Rev. Lett. {\bf 79},
1726 (1997); J. Low. Tem. Phys. {\bf 99}, 443 (1995).

\bibitem{tsuei} See, e.g., C.C. Tsuei and J.P. Kirtley, Rev. Mod.
Phys. {\bf 72}, 969 (2000).

\bibitem{shen1} Z.X. Shen, D.S. Dessau, B.O. Wells, D.M. King,
W.E. Spicer, A.J. Arko, D. Marshall, L.W. Lombardo, A. Kapitulnik,
P. Dickinson, S. Doniach, J. DiCarlo, T. Loeser, and C.H. Park,
Phys. Rev. Lett. {\bf 70}, 1553 (1993); H. Ding, M.R. Norman, J.C.
Campuzano, M. Randeria, A.F. Bellman, T. Yokoya, T. Takahashi, T.
Mochiku, and K. Kadowaki, Phys. Rev. B{\bf 54}, R9678 (1996).

\bibitem{mesot} J. Mesot, M.R. Norman, H. Ding, M. Randeria, J.C.
Campuzano, A. Paramekanti, H.M. Fretwell, A. Kaminski, T.
Takeuchi, T. Yokoya, T. Sato, T. Takahashi, T. Mochiku, and K.
Kadowaki, Phys. Rev. Lett. {\bf 83}, 840 (1999); S.V. Borisenko,
A.A. Kordyuk, T.K. Kim, S. Legner, K.A. Nenkov, M. Knupfer, M.S.
Golden, J. Fink, H. Berger, and R. Follath, Phys. Rev. B{\bf 66},
140509 (2002).

\bibitem{wen} H.H. Wen, H.P. Yang, S.L. Li, X.H. Zeng, A.A.
Soukiassian, W.D. Si, and X.X. Xi, Europhys. Lett. {\bf 64}, 790
(2003).

\bibitem{eliashberg} G.M. Eliashberg, Sov. Phys. JETP {\bf 11},
696 (1960); D.J. Scalapino, J.R. Schrieffer, and J.W. Wilkins,
Phys. Rev. {\bf 148}, 263 (1966).

\bibitem{anderson1} P.W. Anderson, Phys. Rev. Lett. {\bf 67},
2092 (1991); Science {\bf 288}, 480 (2000).

\bibitem{tallon} See, e.g., J.L. Tallon, J.W. Loram, J.R. Cooper,
C. Panagopoulos, and C. Bernhard, Phys. Rev. B {\bf 68}, 180501
(2003).

\bibitem{norman} M.R. Norman, A. Kaminski, J. Mesot, and J.C.
Campuzano, Phys. Rev. B{\bf 63}, 140508 (2001); M.R. Norman, H.
Ding, J.C. Campuzano, T. Takeuchi, M. Randeria, T. Yokoya, T.
Takahashi, T. Mochiku, and K. Kadowaki, Phys. Rev. Lett. {\bf 79},
3506 (1997).

\end{references}
\end{document}